\documentclass[11pt]{article}

\usepackage{tikz}
\usepackage{amssymb,amsmath,amsfonts}
\usepackage{latexsym}
\usepackage[totalwidth=17cm,totalheight=23.8cm]{geometry}

\newcommand\be{\begin{eqnarray}}
\newcommand\ee{\end{eqnarray}}

\newtheorem{proposition}{Proposition}

\begin{document}

\title{Gravitons and a complex of differential operators}
\author{Kirill Krasnov \\ \it{School of Mathematical Sciences, University of Nottingham}\\ \it{University Park, Nottingham, NG7 2RD, UK}}

\date{June 2014}
\maketitle

\begin{abstract}Gravity is now understood to become simple on-shell. We sketch how it becomes simple also off-shell, when reformulated appropriately. Thus, we describe a simple Lagrangian for gravitons that makes use of a certain complex of differential operators. The Lagrangian is constructed analogously to that of Maxwell's theory, just using a different complex. The complex, and therefore also our description of gravitons, makes sense on any half-conformally flat four-dimensional manifold. 
\end{abstract}

\section{Introduction} 

Maxwell's theory of electromagnetism is one of the simplest and most beautiful field theories. Mathematically, much of its beauty and simplicity rests on properties of de Rham complex of differential operators. To state the variational principle that gives rise to field equations it is sufficient to consider only the following part of this complex
\be\label{derham}
\Lambda^0 \stackrel{d}{\longrightarrow} \Lambda^1 \stackrel{d}{\longrightarrow} \Lambda^2 .
\ee
Here $\Lambda^n$ are the spaces of n-forms on a manifold $M$, and $d$ is the operator of exterior derivative. The Lagrangian is then
\be\label{Maxw-L}
{\cal L}\sim (dA)^2, \qquad A\in \Lambda^1,
\ee
where, in view of (\ref{derham}) being a complex, $dA\in \Lambda^2$ is gauge invariant, and the square is computed using the inner product in $\Lambda^2$ that is constructed using the metric on $M$.

A special situation occurs in 4 dimensions. Here the space of two-forms splits $\Lambda^2=\Lambda^+\oplus \Lambda^-$, where $\Lambda^\pm$ are the spaces of self- and anti-self-dual 2-forms, and we get the following resolution of de Rham complex

\bigskip
 \begin{tikzpicture}
\path (0,0) node(a)  {$\Lambda^0$}
         (1.5,0) node(b) {$\Lambda^1$}
         (3.5,1)    node(c)  {$\Lambda^+$}
         (3.5,-1) node(d)  {$\Lambda^-$}
         (5.5,0) node(e) {$\Lambda^3$}
         (7,0) node(f) {$\Lambda^4$};
\draw [->] (a) to node [above] {$d$} (b);
\draw [->] (b) to node [above] {$d_+$} (c);
\draw [->] (b) to node [above] {$d_-$} (d);
\draw [->] (c) to node [above] {$d_+^*$} (e);
\draw [->] (d) to node [above] {$d_-^*$} (e);
\draw [->] (e) to node [above] {$d$} (f);
\end{tikzpicture}
 \bigskip

\noindent The compositions $d_+^* d_+$ and $d_-^* d_-$ are now not equal to zero, but their sum is. We have abused the notation somewhat in denoting the operators from $\Lambda^\pm$ to $\Lambda^3$ by $d_\pm^*$. They become the adjoint of $d_\pm: \Lambda^1 \to \Lambda^\pm$ only after $\Lambda^3$ is identified with $\Lambda^1$ using the metric. It is easy to check that the condition
\be
d_+^* d_+ A = 0 \quad {\rm or} \quad d_-^* d_- A = 0
\ee
is equivalent to the linearised field equation $d^* d A=0$. A general solution of the linearised field equations is then a linear combination of modes $d_+ A=0$ called negative helicity and $d_- A=0$ called positive helicity.

Gravity is as important and ubiquitous as electromagnetism. However, in the description proposed by Einstein gravity is much more complicated than Maxwell's theory. First of all, it is highly non-linear. However, even a linearised description around some (e.g. flat) background is significantly more involved than (\ref{Maxw-L}). In particular, there is no complex of differential operators involved. To be more explicit, the linearisation of the Einstein-Hilbert Lagrangian (around the flat metric) reads 
\be\label{EH-L}
{\cal L}^{(2)}_{\rm EH} = -\frac{1}{2} (\partial_\mu h_{\rho\sigma})^2 + \frac{1}{2} (\partial_\mu h)^2 + (\partial^\mu h_{\mu\nu})^2 + h \partial^\mu \partial^\nu h_{\mu\nu},
\ee
where $h_{\mu\nu}$ is the metric perturbation, and $h:= h^\mu{}_\mu$. This Lagrangian is invariant under $\delta_\xi h_{\mu\nu} = \partial_{(\mu} \xi_{\nu)}$. But, unlike (\ref{Maxw-L}), it cannot be written as a square of a first order differential operator applied to $h_{\mu\nu}$. Indeed, it is clear that such a rewriting of (\ref{EH-L}) is impossible because the Lagrangian does not have a definite sign, even for a Riemannian signature background metric. 

One possible attitude to this complexity is to just admit that gravity is not simple. However, over the last decade we have witnessed the appearance of some striking new results on graviton scattering amplitudes. These turn out to be much simpler than could be expected to result from the Einstein-Hilbert action with its notoriously complicated perturbative expansion, of which (\ref{EH-L}) is the simplest order. Indeed, explicit formulas are now available for $n$-graviton scattering amplitudes in 4 dimensions \cite{Cachazo:2012kg}, as well as in arbitrary number of dimensions \cite{Cachazo:2013hca}. The very fact that such explicit formulas exist is bewildering, given the complexity and number of Feynman diagrams that would need to be summed to produce these amplitudes. Thus, the very existence of closed form answers for the perturbative scattering amplitudes suggests that gravity is much simpler than its Einstein-Hilbert formulation suggests. 

The purpose of this exposition is to advertise the fact that gravity in four dimensions simplifies not only on-shell, but, when reformulated appropriately, also off-shell. Thus, we will show that there exists a Lagrangian description of gravitons that is much simpler than (\ref{EH-L}). The new description is based on a certain complex of differential operators, and the linearised Lagrangian is the direct analog of (\ref{Maxw-L}). This description has been worked out in a series of works \cite{Krasnov:2011pp}-\cite{Fine:2013qta} by this author and collaborators on the so-called pure connection formulation of General Relativity. However, the fact that there is complex of differential operators underlying this description is noted here for the first time. Our description is self-contained; no familiarity with the previous literature is assumed. 

\section{Spinor bundles and related differential operators}

We start by describing two (closely related) infinite diagrams of differential operators. These diagrams house many complexes of differential operators, and the complex of interest for us for applications to gravity will be one of them. However, it is worth to describe things in more generality, because some aspects of the construction become clearer. 

Let us consider a 4-dimensional Riemannian manifold $M$, and let $\nabla$ be the operator of covariant differentiation. Let us also assume that $M$ is a spin manifold so that the two-dimensional spinor bundles $S_\pm$ exist. Then, as is well-known, the tangent and cotangent bundles can be identified with the bundle $S_+\otimes S_-$. A general irreducible spinor bundle is of the form $S_+^k \otimes S_-^n$, where the power denotes the symmetrised tensor product. In particular, ${\rm dim}(S_\pm^k) = k+1$. Any tensor or spinor field is then an object in a bundle of this type, or in a direct sum of such bundles. Let us introduce a convenient notion of the {\it spin} of a spinor bundle:
\be\label{spin}
J( S_+^k\otimes S_-^n) := (k+n)/2.
\ee
This notion of spin coincides with the usual physics definition. Indeed, the spin of a field is determined by its transformation properties with respect to the group of spatial rotations, after a space plus time decomposition of the manifold is performed. This decomposition defines an ${\rm SO}(3)\subset{\rm SO}(4)$ that fixes the "time" direction. Under this group both spinors $S_\pm$ transform as the spin half representations of ${\rm SO}(3)$, and so (\ref{spin}) is indeed the total spin. 

Let us now introduce two types of first order differential operators. Both are constructed from the operator of covariant derivative $\nabla_\mu$, that in spinor notations becomes $\nabla_{AA'}$. We denote objects in $S_+$ as those with unprimed spinor indices, while the primed index objects are in $S_-$. We raise and lower spinor indices with the spinor metrics $\epsilon^{AB}, \epsilon_{AB}$, and similarly for the primed indices. As usual with spinors, one has to be careful about the order of the spinor indices, as raising-lowering a pair produces a minus sign. A consistent set of rules for this can be developed, but we have not tried to do it here in order not to overburden this paper with conventions. An interested reader will easily fill in the details. 

The operator of first type, which we denote by $d$, increases the spin of a field by one unit $d: J\to J+1$ and acts according to
\be
d: S_+^k \otimes S_-^n \to S_+^{k+1} \otimes S_-^{n+1}, \qquad \psi^{A_1\ldots A_k A_1'\ldots A_n'} \to (k+1)(n+1)\nabla^{(A_{k+1}(A'_{n+1}} \psi^{A_1'\ldots A_n') A_1\ldots A_k)}.
\ee
In other words, one first acts by the operator of covariant derivative, and then symmetrises all the resulting spinor indices. The normalisation is included for later convenience. Note that the operator $d$ on functions $d: {\mathcal C}^\infty(M)\to S_+\otimes S_-$ is just the usual (exterior) derivative mapping functions to one-forms. This justifies using the notation $d$ for this operator acting on other spinor bundles as well. 

The other operator does not change the spin $\delta: J\to J$, but instead flips one unit of spin from the space of the primed spinor to the unprimed ones
\be
\delta: S_+^k \otimes S_-^n \to S_+^{k+1} \otimes S_-^{n-1}, \qquad \psi^{A_1\ldots A_k A_1'\ldots A_n'} \to (k+1) \nabla^{(A_{k+1}}_{A'} \psi^{A_1\ldots A_k)A_1'\ldots A_{n-1}' A'}.
\ee
Note that the action of $\delta$ on the fundamental spinor bundle $S_-$ is just that of the Dirac operator $\delta: S_-\to S_+$, so $\delta$ is a version of the Dirac operator.

We will also need the adjoint of $\delta$. To compute this, we need an inner product on all our spinor bundles. A natural definition is to take a product of two spinors using the spinor metric, and integrate over $M$, including the normalisation factor $(k! n!)^{-1}$. With this definition of the product, the adjoint of $\delta$ is given by
\be
\delta^*: S_+^k \otimes S_-^n \to S_+^{k-1} \otimes S_-^{n+1}, \qquad \psi^{A_1\ldots A_k A_1'\ldots A_n'} \to (n+1) \nabla^{A(A_{n+1}'} \psi^{A_1'\ldots A_{n}') A_1 \ldots A_{k-1}}{}_A.
\ee
This is also the Dirac operator, with the action on the fundamental bundle $S_+$ being $\delta^*: S_+\to S_-$. 

Now, applying the operators $d,\delta,\delta^*$ to the space ${\mathcal C}^\infty(M)$ we obtain the diagram shown in Fig.~1. It consists of spaces as nodes and differential operators as maps between nodes. What is shown in Fig.~1 is only the first 3 rows of the diagram, but the pattern can be continued and the diagram contains an infinite number of rows. We have only drawn it up to the spin 2 line, which is what is relevant for us here.

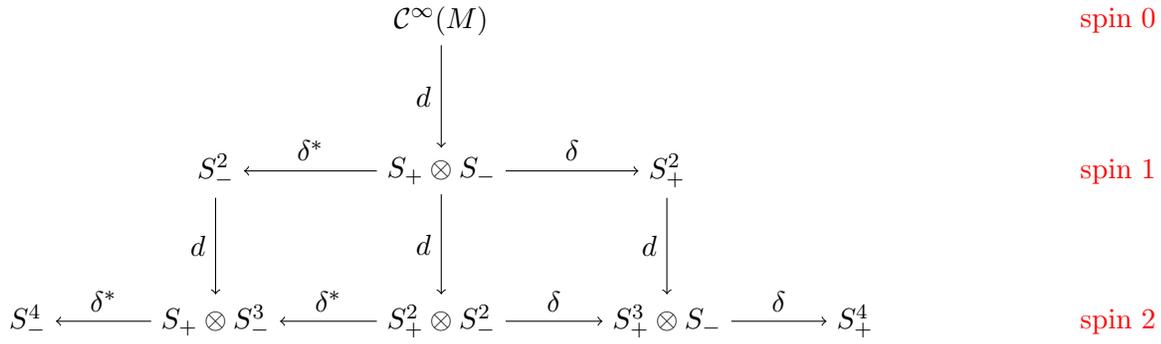
\begin{figure}\label{fig-1}
 \begin{tikzpicture}
\path (0,0) node(a)  {${\mathcal C}^\infty(M)$}
          (0,-2) node(b)  {$S_+\otimes S_-$}
          (3,-2) node(c)  {$S_+^2$}
          (0,-4) node(d)  {$S_+^2\otimes S_-^2$}
          (3,-4) node(e)  {$S_+^3\otimes S_-$}
          (5.5,-4) node(f)  {$S_+^4$}
          (-3,-2) node(s) {$S_-^2$}
          (-3,-4) node(t) {$S_+\otimes S_-^3$}
          (-5.5,-4) node(u) {$S_-^4$}
          (9,0) node(p) [red] {spin 0}
          (9,-2) node(q) [red] {spin 1}
          (9,-4) node(r) [red] {spin 2};
\draw [->] (a) to node [left] {$d$} (b);
\draw [->] (b) to node [above] {$\delta$} (c);
\draw [->] (b) to node [left] {$d$} (d);
\draw [->] (c) to node [left] {$d$} (e);
\draw [->] (d) to node [above] {$\delta$} (e);
\draw [->] (e) to node [above] {$\delta$} (f);
\draw [->] (s) to node [left] {$d$} (t);
\draw [->] (b) to node [above] {$\delta^*$} (s);
\draw [->] (d) to node [above] {$\delta^*$} (t);
\draw [->] (t) to node [above] {$\delta^*$} (u);
\end{tikzpicture}
\caption{The integer spin diagram}
 \end{figure}

It is important to emphasise that, in general, this diagram is not a complex, i.e. the composition of the operators does not give zero. Neither it is in general commutative. However, with some additional assumptions about the curvature the diagram receives the following properties:
\begin{proposition} The diagram commutes in the direction down and right if and only if the self-dual part of the Weyl curvature $W^+$ is zero, and the metric is Einstein. In other words 
\be
(d\delta - \delta d)(S_+^k\otimes S_-^n)=0 \qquad \Leftrightarrow \qquad W^+=0, \quad ({\rm Ricci}) \sim ({\rm metric}). 
\ee
Moreover, on purely chiral spaces $S_+^k$ the $d\delta$ in the above equation is absent, and the commutativity property reduces to 
\be
\delta d (S_+^k)=0 \qquad \Leftrightarrow \qquad W^+=0.
\ee
In this case it is not necessary for the metric to be Einstein, it is enough that $W^+=0$. On the other hand, on purely chiral spaces of the other chirality $S_-^n$ it is enough for the metric to be Einstein for the diagram to commute 
\be
(d\delta - \delta d)(S_-^n)=0  \qquad \Leftrightarrow \qquad ({\rm Ricci}) \sim ({\rm metric}). 
\ee
\end{proposition}
A similar property set of properties holds if one reads the diagram in the direction from right to left. Here one just has to use the adjoint operator $\delta^*$ instead. 
\begin{proposition} The diagram commutes in the direction down and left if and only if the self-dual part of the Weyl curvature $W^-$ is zero, and the metric is Einstein. In other words 
\be
(d\delta^* - \delta^* d)(S_+^k\otimes S_-^n)=0 \qquad \Leftrightarrow \qquad W^-=0, \quad ({\rm Ricci}) \sim ({\rm metric}). 
\ee
Moreover, on purely chiral spaces $S_-^n$ the $d\delta^*$ in the above equation is absent, and the commutativity property reduces to 
\be
\delta^* d (S_-^n)=0 \qquad \Leftrightarrow \qquad W^-=0.
\ee
In this case it is not necessary for the metric to be Einstein, it is enough that $W^-=0$. On the other hand, on purely chiral spaces of the other chirality $S_+^k$ it is enough for the metric to be Einstein for the diagram to commute 
\be
(d\delta^* - \delta^* d)(S_+^k)=0  \qquad \Leftrightarrow \qquad ({\rm Ricci}) \sim ({\rm metric}). 
\ee
\end{proposition}
In all cases the statements follow from the fact that the left-hand-sides of the above equations compute some components of the curvature. When the relevant components are zero, the result is zero. 

These two propositions have a number of interesting corollaries. First, the whole infinite diagram shown above is commutative on a conformally flat space, such as $S^4$. A more interesting corollary is that the part of the diagram shown above, i.e. only up to the spin 2 spinor bundles, has the property that its right-hand-side containing spaces $k\geq n$ is commutative when $W^+=0$ and the metric is Einstein, i.e. on what is called the gravitational instantons. 

Before we consider the spin 2 part of the above diagram in more detail, let us point out that there is another infinite diagram of the same sort, obtained by starting from a spin $1/2$ bundle instead. Indeed, starting with e.g. $S_+$ and applying $d, \delta, \delta^*$ we obtain the diagram shown in Fig.~2.  It is continued to the next row of spin $5/2$ and so on. 
\begin{figure}
 \begin{tikzpicture}
\path (0,0) node(a)  {$S_+$}
          (-3,0) node(b)  {$S_-$}
          (0,-2) node(c)  {$S_+^2\otimes S_-$}
          (-3,-2) node(d)  {$S_+\otimes S_-^2$}
          (3,-2) node(e)  {$S_+^3$}
          (-6,-2) node(f)  {$S_-^3$}
          (9,0) node(p) [red] {spin 1/2}
          (9,-2) node(q) [red] {spin 3/2};
\draw [->] (a) to node [left] {$d$} (c);
\draw [->] (c) to node [above] {$\delta$} (e);
\draw [->] (b) to node [left] {$d$} (d);
\draw [->] (a) to node [above] {$\delta^*$} (b);
\draw [->] (d) to node [above] {$\delta^*$} (f);
\draw [->] (c) to node [above] {$\delta^*$} (d);
\end{tikzpicture}
\caption{The half-integer spin diagram}
\end{figure}
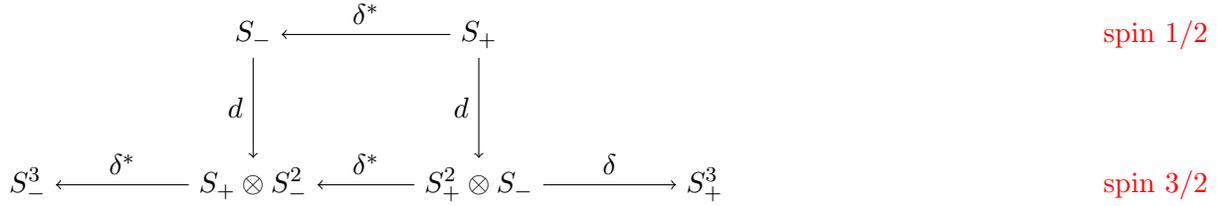 
  
The two propositions above apply also in this case. Thus, if we concentrate just on the part of the diagram shown in Fig.~2, then the box is commutative whenever the metric is Einstein. We also have two complexes
 \be
 \delta d(S_+)=0 \qquad \Leftrightarrow \qquad W^+=0, \\ \nonumber 
 \delta^* d(S_-)=0 \qquad \Leftrightarrow \qquad W^-=0.
 \ee
 The fact that the box in the diagram of Fig.~2 commutes iff the metric is Einstein was pointed out a long time ago in \cite{Julia:1982ks}. Let us give a quick proof of this in our notations, which also illustrates how the more general statements of Propositions 1,2 are deduced. The fact that the diagram commutes implies that 
\be
 S_+ \stackrel{d+\delta^*}{\longrightarrow} (S_+^2 \otimes S_-)\oplus(S_-)  \stackrel{\delta^*-d}{\longrightarrow} S_+ \otimes S_-^2
 \ee 
 constructed by following the diagram from the top right corner to the bottom left is a complex. However, noting that $(S_+^2 \otimes S_-)\oplus(S_-)=S_+\otimes S_+\otimes S_-$ we can rewrite the first map as
 \be
 (d+\delta^*): S_+ \to S_+\otimes S_+\otimes S_-, \qquad \psi^A \to \nabla^{AA'} \psi^B.
 \ee
 The second map is then
 \be
 (\delta^*-d): S_+\otimes S_+\otimes S_- \to S_+ \otimes S_-^2, \qquad \psi^{ABA'}\to \nabla^{E(A'} \psi_E{}^{B')A},
 \ee
 and the fact that the composition is zero iff the metric is Einstein follows from
 \be
 \nabla^{E(A'} \nabla_E{}^{B')} \psi^A = R^{A'B' AB} \psi_B,
 \ee
 where $R^{A'B' AB}\in S_+^2\otimes S_-^2$ is the tracefree part of the Ricci curvature. 
 
 In article \cite{Julia:1982ks} the complex property is noted for the composition in the different direction
 \be
 S_+\otimes S_-^2 \stackrel{\delta-d^*}{\longrightarrow} S_+\otimes S_+\otimes S_- \stackrel{d^*+\delta}{\longrightarrow} S_+.
 \ee
The result of the first map being zero is then the field equation for a spin $3/2$ particle. The composition of the two maps being zero is the integrability property of the field equations, which thus holds iff the metric is Einstein. 

Having related the commutativity of the diagram in Fig.~2 to the findings of \cite{Julia:1982ks} we can say that our Propositions 1,2, generalise the observation of this reference to a setting of more general spinor bundles.
 
 Our last comment in this section is that the Propositions 1,2 are not at all difficult to prove. This author is informed that they are known (albeit in a slightly different form) to experts in this area. However, they seem to be part of the folklore that mathematicians with this sort of background learn, and no reference seems to exist, at least not any references describing things in this generality. We hope that spelling these facts out will benefit the community. 
 
 The related facts that are available in the literature are as follows. The fact that $\delta d (S_+^k)=0$ whenever $W^+=0$ is noted in Besse \cite{Besse:1987pua}, see Proposition 13.27. The other relevant reference we are aware of is the already cited paper \cite{Julia:1982ks} by Julia. 
 
 In the next section we will list the complexes that can be obtained from the above diagram for the case of spin 2. 
 
\section{Spin two}

Before we discuss applications to gravity, let us start by listing all complexes that can be read off the diagram shown in the previous section. We concentrate only on the right half of the diagram, as the other half is similar. We obtain
\be\label{c1}
{\mathcal C}^\infty \stackrel{d}{\longrightarrow} S_+\otimes S_- \stackrel{\delta}{\longrightarrow} S_+^2, \\  \label{c2}
S_+^2 \stackrel{d}{\longrightarrow} S_+^3\otimes S_- \stackrel{\delta}{\longrightarrow} S_+^4 \qquad \Leftrightarrow \qquad W^+=0.
\ee
These directly follow from the Proposition 1 of the previous section. Note that the first complex here is just a part of the de Rham complex resolution that we discussed in the Introduction. Indeed, $\Lambda^+\sim S_+^2$, and so (\ref{c1}) directly follows from de Rham complex. Thus, the second line can be thought of as a generalisation of de Rham complex. Note, however, that one needs the manifold to be half- conformally flat. It is this complex that we will use for our description of gravitons below. 

The other two complexes are obtained by using the commutativity of the diagram in Fig.~1. They read
\be\label{c3}
S_+\otimes S_- \stackrel{d+\delta}{\longrightarrow} S_+^2\otimes S_-\otimes S_-  \stackrel{\delta-d}{\longrightarrow} S_+^3\otimes S_- \qquad &\Leftrightarrow& \qquad W^+=0, \quad ({\rm Ricci}) \sim ({\rm metric}), \\ \label{c4}
S_+^2\stackrel{d+\delta^*}{\longrightarrow} S_+^2\otimes S_+\otimes S_-  \stackrel{\delta^*-d}{\longrightarrow} S_+^2\otimes S_-^2 \qquad &\Leftrightarrow& \qquad ({\rm Ricci}) \sim ({\rm metric}).
\ee
The notation here is self-explanatory, for example the first map in (\ref{c3}) is given by $d+\delta: \psi^{AA'} \to \nabla^{A'(A} \psi^{B)B'} \in S_+^2\otimes S_-\otimes S_-$. Decomposing this space into irreducibles we get the two spaces $S_+^2$ and $S_+^2\otimes S_-^2$ that appear when one follows the diagram in Fig.~1 from $S_+\otimes S_-$ to the right and to the bottom. The second map in (\ref{c3}) acts by converting the primed index $A'$ to an unprimed index, and then symmetrising all resulting unprimed indices. The complexes (\ref{c3}),(\ref{c4}) appear to have been unnoticed before. The complex (\ref{c4}) is particularly interesting, for it only requires the space to be Einstein, with no further assumptions on the curvature.

\section{Gravitons}

Both of the complexes (\ref{c1}), (\ref{c2}) describe a system with two degrees of freedom, which can be confirmed by subtracting the dimension of the first space (the space where gauge transformation parameters take value) from the dimension of the last space. Equivalently, the number of degrees of freedom is the dimension of the space of fields (the middle space) minus twice the dimension of the gauge parameter space. For (\ref{c1}), (\ref{c2}) these dimensions read
\be\label{c1-d}
1 \stackrel{d}{\longrightarrow} 4 \stackrel{\delta}{\longrightarrow} 3, \\  \label{c2-d}
3 \stackrel{d}{\longrightarrow} 8 \stackrel{\delta}{\longrightarrow} 5,
\ee
and so each of them is capable of describing two propagating degrees of freedom, as is appropriate for a massless particle in 4 dimensions. The first one involves spin one fields, and is thus appropriate for the description of electromagnetism. The second one is about spin two fields, and thus can be suspected to have something to do with gravity. We thus write down the following Lagrangian
\be\label{L-gr}
{\cal L}^{(2)} \sim (\delta a)^2, \quad a\in S_+^3\otimes S_-.
\ee
In view of (\ref{c2}), it is gauge-invariant on any half-conformally flat space. Performing a simple canonical analysis one finds \cite{Delfino:2012zy} that this Lagrangian indeed describes massless spin 2 particles -- gravitons. Thus, one finds that after the constraints corresponding to gauge rotations are imposed, and the residual gauge freedom is fixed, the phase space reduces to a pair of symmetric traceless transverse spatial tensors. The arising reduced Hamiltonian is also the standard one for gravitons. 

The field equations that follow from (\ref{L-gr}) are
\be
\delta^* \delta a = 0.
\ee
As in the case of Maxwell theory, a particularly simple set of solutions of these equations is given by fields satisfying the first order equation $\delta a=0$. These describe fields of negative helicity. A description of the positive helicity fields is more complicated, see \cite{Delfino:2012zy} for details. 

\section{Discussion}

Gravitons are self-interacting, and so the natural question is if there is a non-linear completion of the above linearised description. The answer is in the affirmative. In fact, the linearised description given above was obtained by analysing the pure connection formulation of General Relativity \cite{Krasnov:2011pp}. The linearisation around an instanton background, as well as the first orders in the perturbative expansion are described in \cite{Delfino:2012aj}. One notable fact about the resulting interaction terms is that they are much more compact than in the usual metric description. 

For completeness, let us sketch how the above linearised description arises. In the pure connection formulation gravity is described as a dynamical theory of an ${\rm SO}(3)$ connection. On an appropriate background the internal index of the connection perturbation one-form becomes identified with the space $S_+^2$. Thus, the connection perturbation takes values in the space $S_+^2\otimes S_+\otimes S_-$. This decomposes as $S_+\otimes S_-$ and $S_+^3\otimes S_-$. The linearised Lagrangian is then just a function of the $S_+^3\otimes S_-$ component, and is given by (\ref{L-gr}). In other words, the linearised Lagrangian is invariant under shifts of the $S_+\otimes S_-$ component. This is how diffeomorphism invariance enters the story. However, unlike in the metric formulation, the action of the linearised diffeomorphisms on the connections is very simple, and the diffeomoprhism part of the connection perturbation can be projected out. In other words, the natural gauge-fixing condition for the diffeomorhisms in the language of  connections is algebraic, not differential. Such a gauge-fixing condition does not make the corresponding components propagating. This results in just the gravitational perturbation theory with a single irreducible representation $S_+^3\otimes S_-$ of the Lorentz group propagating, see \cite{Delfino:2012aj} for more details. 

The diagram (\ref{c2}) only becomes a complex on a half-conformally flat space. Thus, our description of gravitons only works on such spaces. It is an open problem to find whether the linearised dynamics of gravitons can similarly be reduced to some complex of differential operators on any Einstein manifold. This would utilise some complex that only imposes the constraint that the tracefree part of Ricci tensor is zero, similar to (\ref{c4}). However, the complex required should also describe the correct number of degrees of freedom, which is not the case with (\ref{c4}). This interesting and important problem is the subject of current investigation. 

\section*{Acknowledgements} The author was supported by an ERC Starting Grant 277570-DIGT, and is grateful to Joel Fine, Laurent Freidel, Lionel Mason as well as Nigel Hitchin for discussions.

\end{document}